\documentclass[aps, prl, 10pt, superscriptaddress, twocolumn]{revtex4-1}
	\usepackage[bookmarks=false,linkcolor=blue,urlcolor=blue,colorlinks,citecolor=blue]{hyperref}

	\usepackage{graphicx}
	\usepackage{color}
	\usepackage{amsmath}
	\usepackage[normalem]{ulem}
	\usepackage[toc,page]{appendix}
	\usepackage{amsfonts}
	\usepackage{amssymb}

\newcommand{\be}{\begin{equation}}
\newcommand{\ee}{\end{equation}}
\newcommand{\p}{\partial}

\newcommand{\la}{\label}
\newcommand{\bea}{\begin{eqnarray}}
\newcommand{\eea}{\end{eqnarray}}
\newcommand{\SUM}{\sum_{i=1}^{\mathrm{N}_{\rm el}}}

\newcommand{\bz}{\{ z_i\}}
\newcommand{\bk}{\mathbf{k}}
\newcommand{\bq}{\mathbf{q}}
\newcommand{\cC}{{\cal C}}
\newcommand{\cO}{{\cal O}}
\newcommand{\cT}{{\cal T}}
\newcommand{\cV}{{\cal V}}
\newcommand{\half}{{\frac12}}

\def\sst{\scriptscriptstyle}

\definecolor{cardinal}{rgb}{0.6,0,0}
\definecolor{darkgreen}{rgb}{0,0.4,0}
\definecolor{golden}{rgb}{0.92, 0.7, 0}
\definecolor{midnight}{rgb}{0, 0, 0.5}
\definecolor{darkblue}{rgb}{0, 0, 0.7}

\begin{document}

\title{\begin{flushright}\vspace{-1in}
			\mbox{\normalsize  EFI-XX-XX}
		\end{flushright}
	Collective excitations at filling factor $5/2$: The view from superspace
	 \vskip 20pt
	 }

\author{Andrey Gromov}
\affiliation{Brown Theoretical Physics Center and Department of Physics,
Brown University, Providence, Rhode Island 02912, USA}
\author{Emil J. Martinec}
\affiliation{Kadanoff Center for Theoretical Physics, University of Chicago, Chicago, Illinois 60637}
\affiliation{Enrico Fermi Institute, University of Chicago, Chicago, Illinois 60637, U.S.A}
\author{Shinsei Ryu}
\affiliation{Kadanoff Center for Theoretical Physics, University of Chicago, Chicago, Illinois 60637}
\affiliation{James Franck Institute, University of Chicago, Chicago, Illinois 60637, U.S.A}

\date{\today}

%%%%%%%%%%%%%%%%%%%%%%%%%%%%%%%%%%%%%%%%%%%%%%%%%%%%%%%%%%%%%%%%%%%%%%%%%%%%%%%%%%%%%%%%%%%%%%%
\begin{abstract}
We present a microscopic theory of the neutral collective modes supported by the non-Abelian fractional quantum Hall states at filling factor $5/2$. The theory is formulated in terms of the trial states describing the Girvin-MacDonald-Platzman (GMP) mode and its fermionic counterpart. These modes are superpartners of each other in a concrete sense, which we elucidate.
\end{abstract}

%%%%%%%%%%%%%%%%%%%%%%%%%%%%%%%%%%%%%%%%%%%%%%%%%%%%%%%%%%%%%%%%%%%%%%%%%%%%%%%%%%%%%%%%%%%%%%%

\maketitle

%%%%%%%%%%%%%%%%%%%%%%%%%%%%%%%%%%%%%%%%%%%%%%%%%%%%%%%%%%%%%%%%%%%%%%%%%%%%%%%%%%%%%%%%%%%%%%%

%%%%%%%%%%%%%%%%%%%%
\paragraph{Introduction.---} 
%%%%%%%%%%%%%%%%%%%%
The experimental discovery of fractional quantum Hall states has stimulated the development of the first-quantized approach to strongly interacting many-body problems. The model state introduced by Laughlin has been extremely successful in describing qualitative features of fractional quantum Hall phases as quantum fluids with fractionally charged excitations \cite{laughlin1983anomalous}.  Shortly after Laughlin's work Girvin, MacDonald and Platzman (GMP) developed a theory of the neutral collective excitation (known as the magneto-roton) supported by the Laughlin phase \cite{girvin1986magneto}. This collective mode has been observed in the Raman scattering experiments \cite{pinczuk1993observation, kukushkin2009dispersion}.  GMP estimated the value of various gaps and computed the entire dispersion curve of the mode. It was later found \cite{Papic-SMA} that the dispersion curve is very accurate at long wavelengths, but breaks down at the magnetic length scale. The GMP theory relies on the fact that in a strong magnetic field it is possible to neglect the transitions to higher Landau levels and work in the restricted Hilbert space of lowest Landau level (LLL) states. After the LLL restriction, the density operators no longer commute with each other and instead form the algebra of area-preserving diffeomorphisms (APD), $W_\infty$. This algebra plays a central role in the GMP computations of the dispersion of the mode. 

Moore and Read (MR) introduced a trial state \cite{moore1991nonabelions} to
describe qualitative properties of a quantum Hall plateau that forms at the
filling fraction $\nu = \frac{5}{2}$ \cite{willett1987observation}. This
construction predicts that the state has non-Abelian topological order and the
fractionally charged excitations with electric charge $\frac{e}{4}$.  Recently,
there has been a resurgence of interest in the $\frac{5}{2}$ state after the
first measurement of the thermal Hall conductance
\cite{banerjee2017observation}, which suggests the non-Abelian nature of the
state, and the observation of the plateau at $\nu = \frac{5}{2}$ in bi-layer
graphene \cite{li2017even}.  Regarding collective modes, the construction of GMP
applies equally well to the Moore-Read state and the corresponding magneto-roton
mode is expected.  Greiter, Wen and Wilczek (GWW) \cite{greiter1991paired}
proposed that due to the paired nature of the MR state, there should be another
collective mode corresponding to breaking a Cooper pair. Subsequent numerical
work has shown that there are indeed two collective modes present in the
spectrum of a Hamiltonian, that supports the MR state as its groundstate
\cite{bonderson2011numerical, moller2011neutral, yang2012model}.   Ref.\ \cite{yang2012model} has used Jack polynomials to construct a series of trial states that accurately describe the fermionic mode. However, a simple and intuitive construction of the trial state a l\'a GMP has been lacking.

In this Letter we use an auxiliary superspace formalism to construct the GMP and GWW trial states in a uniform fashion.  We compute the gap function of the neutral fermion mode. Within this framework it can be made precise that the two modes are superpartners of one another.  Additionally, we make a connection between the conformal field theory construction of the trial states and the collective neutral modes.

%%%%%%%%%%
\paragraph{The GMP mode.---}
%%%%%%%%%%

We start with a brief review of the single mode approximation developed by GMP. We are interested in approximating an excited collective mode of a $2$-body Hamiltonian, projected to the lowest Landau level
\be\la{eq:Ham}
H = \int d^2 \bq V_\bq \bar \rho_{-\bq}\bar \rho_{\bq}\,,
\ee
where $V_\bq$ is the Fourier transform of the interaction potential and $\bar \rho_\mathbf{k}$ is the (normal ordered) density operator projected to the lowest Landau level \cite{girvin1984formalism}
\be\la{eq:pDen}
\bar \rho_{\mathbf{k}} =  \SUM e^{- i \bar k \p_i} e^{- \frac{i}{2}  k  z_i} \,, 
\ee
where we have set the magnetic length $\ell$ to unity.
Projected density operators satisfy the $W_\infty$ algebra 
\be\la{eq:GMP}
[\bar \rho_\bk, \bar \rho_\bq] =  f(\bk,\bq)\bar \rho_{\bk+\bq}\,,\quad f(\bk,\bq) = 2i ~ e^{\frac{\mathbf{k} \cdot \mathbf{q}}{2} }~ \sin \left( \frac{ \mathbf{k} \times \mathbf{q}}{2} \right)\,.
\ee
The GMP mode is a projected density wave
\be
|\bk\rangle = \bar \rho_\bk |0\rangle\,,
\ee
where $|0\rangle$ is the exact groundstate of $H$. The particular form of the groundstate $|0\rangle$ depends on $H$, but plays no role in the construction.  The energy of the GMP mode is given by \cite{girvin1986magneto}
\be
\Delta_{GMP}(\bk) = \frac{\langle \bk| H |\bk \rangle}{\langle \bk| \bk \rangle} = \frac{\bar f(\bk)}{\bar s(\bk)}\,,
\ee
where $\bar f(\bk)$ and $\bar s(\bk)$ are the oscillator strength and the projected static structure factor (SSF) respectively
\be
\label{densityalg}
\bar f(\bk) = \frac{1}{2} \langle 0| [\bar \rho_{-\bk},[H, \bar \rho_{\bk}]] |0\rangle\,, \quad \bar s(\bk) = \langle 0| \bar \rho_{-\bk} \bar \rho_{\bk} |0\rangle\,.
\ee
Quite a lot is known about the projected SSF for various quantum Hall states
\cite{girvin1986magneto,CLW,can2014geometry, Gromov-galilean, golkar2016spectral, nguyen2014lowest, Nguyen-PHD, Haldane:2009p15460, haldane2011geometrical, haldane2011self, nguyen2018algebraic, read2011hall, lee2018pomeranchuk}. For chiral trial states the small momentum expansion of the SSF takes the form
\be
\bar s(\bk) = s_4 |k|^4 + s_6 |k|^6 + \ldots\,, 
\ee
where $s_4 = \frac{|\mathcal{S} - 1|}{8}$ is determined by the Wen-Zee shift \cite{WenZeeShiftPaper} and $s_6$ is determined by the shift and central charge \cite{Nguyen-PHD, GCYFA}. Under more general conditions the long wave expansion of the SSF still starts with $|k|^4$. The oscillator strength depends on the microscopic details, but given the general structure of SSF, the long wave expansion must take the form $\bar f(\bk) = f_4|k|^4 + \ldots$ in order to ensure the finite value of $\Delta_{GMP}(\bk=0)$. The general expression for $\bar f(\bk)$ is
\bea
&&\bar f(\bk) = 4\int d^2 \bq~ V_\bq \left(\sin \left( \frac{\bk\times \bq}{2}\right)\right)^2  F(\bq,\bk)\,, 
\\
&&F(\bq,\bk) =  e^{\mathbf k\cdot \mathbf q}\bar s(\bq+\bk)  +  e^{-\frac{|\bk|^2}{2}}\bar s(\bq) \,.
\nonumber 
\eea
Thus, the gap function $\Delta(\mathbf{k})$ is determined by the projected SSF and the interaction.

%%%%%%%%%%
\paragraph{Trial states on a superplane.---}
%%%%%%%%%%

In this section we will introduce the auxiliary superspace construction --- the main technical tool employed in this Letter. We will make use of the quantum Hall problem formulated on a superplane~\cite{hasebe2005supersymmetric}, which unifies bosonic Laughlin and fermionic Moore-Read states into a single object~\cite{hasebe2008unification}. 

The superplane $\mathbb R^{2|2}$ is characterized by two sets of coordinates -- bosonic $z, \bar z$, and fermionic (\emph{i.e.} anticommuting) $\theta, \bar \theta$.  Every electron ``living'' on a superplane is characterized by such a pair of (holomorphic) coordinates: $(z_i, \theta_i)$.  Many-body super-Laughlin state of electrons is given by
\be\la{eq:sL}
\Psi_{sL} = \prod_{i<j} (z_i-z_j - \theta_i\theta_j)^2 e^{- \SUM \frac{|z_i|^2}{4}}\,. 
\ee
The quickest way to obtain this state is to generalize the Moore-Read
construction to the superplane. To do so we consider the $U(1)_2 \times \it {\it
  Ising}$ CFT and define a superfield
\be
\label{eq:charge superfield}
\Phi(z,\theta) = e^{i\sqrt{2} (\phi(z)+\theta\psi(z))} = e^{i\sqrt{2}\phi(z)} \bigl(1 + i\sqrt{2}\theta \psi(z) \bigr)\,,
\ee 
where $e^{i\sqrt{2}\phi(z)}$ is the charge $1$ bosonic vertex operator and $\psi(z)$ is the Majorana fermion. Then the super-Laughlin state is obtained as a superconformal block
\be
\label{superLaugh}
\Psi_{sL} = \left\langle \prod_{i=1}^{\mathrm{N}_{\rm el}} \Phi(z_i,\theta_i) \mathcal Q  \right\rangle\,,
\ee 
where $\mathcal Q$ is the background charge operator that ensures non-vanishing of the correlation function \cite{moore1991nonabelions}
\be
\mathcal Q = \exp\Bigl[{-i\int d^2z \bigl( \rho_0 \sqrt{2} \phi(z)} + \lambda(z,\bar z) \psi(z) \bigr)\Bigr]\,.
\ee
Here $\lambda$ is a source for $\psi$ that can be chosen to supply the proper fermion parity in correlators. When $\mathrm{N}_{\rm el}$ is an even number the construction reduces exactly to the state \eqref{superLaugh}. When $\mathrm{N}_{\rm el}$ is odd we choose $\lambda(z,\bar z)$ to be supported on a line that encircles all of the electrons.

We will concentrate on the even $\mathrm{N}_{\rm el}$ case first and discuss the
odd case later.
Remarkably, the Moore-Read state is the highest component
\footnote{Here the ``highest component'' is meant in the superspace sense. It is a function of $z$, that is multiplied by the largest possible number of $\theta$s.} of the super-Laughlin state~\eqref{superLaugh}
\be
\Psi_{\it MR} = \int \prod_{i=1}^{\mathrm{N}_{\rm el}}d\theta_{i} \Psi_{sL} = \mathrm{Pf}\left(\frac{1}{z_i-z_j}\right)\Delta^2e^{-\SUM \frac{|z_i|^2}{4}}\,,
\ee
where $\Delta = \prod_{i<j} (z_i-z_j)$ is the Vandermonde determinant and $\mathrm{Pf}(M_{ij})$ is the Pfaffian of the matrix $M_{ij}$. 
This identity can be seen by Taylor expanding \eqref{eq:sL} in $\theta$.  The integration over all $\theta_i$ picks out the set of terms linear in all $\theta_i$, yielding the Pfaffian factor which automatically comes out antisymmetrized. We are led to a natural strategy. Since the Moore-Read state is much simpler \emph{before} taking the $\theta$-integral, then it is easier to perform various computations on the superplane first and integrate over $\theta$ in the end. Due to the anti-commuting nature of $\theta$-variables we will always obtain fully antisymmetric wavefunctions.  

%%%%%%%%%%
\paragraph{Density on a superplane.---}
%%%%%%%%%%

We turn to the neutral collective modes. It is natural to consider the density operator on a superplane (or \emph{superdensity} operator) defined as
\be
\rho(z,\theta) = \SUM \delta(z-z_i)\delta(\theta-\theta_i)\,.
\ee
Using the fact that Grassmann $\delta$-function of a complex variable $\theta$ is quadratic
\be
\delta(\theta-\theta_i) = (\theta - \theta_i)(\bar \theta - \bar \theta_i)\,,
\ee
we can expand the superdensity $\rho(z,\theta)$ as
\be
\rho(z,\theta) = r(z) + \bar \theta\eta(z) + \theta \eta^\star(z) + \theta \bar \theta \rho(z)\,.
\ee
Here we introduced the following operators
\bea
\eta(z) =\SUM &\theta_i& \delta(z-z_i)\,,\quad \eta^\star(z) = \SUM \bar\theta_i \delta(z-z_i)\,,
\\
&& r(z) = \SUM \theta_i \bar \theta_i\delta(z-z_i)\,.
\eea
The operators $\bar \eta$ and $\bar \eta^\star$ should be viewed as spin-$1/2$ superpartners of the bosonic density $\rho$.

We take the Fourier transform with respect to both even and odd coordinates.
The LLL projected superdensity operator is then given by
\bea\la{eq:sDensity}
&&\bar \rho_{\bk,\varkappa} = \SUM e^{-i\bar k \partial_i} e^{-\frac{i}{2} k z_i}    e^{-\frac{i}{2} \varkappa \theta_i}e^{-\frac{i}{2} \bar \varkappa \bar\theta_i}
\\
 &&= \bar \rho_\bk -\frac{i}{2}\varkappa\bar\eta^\star_\bk -\frac{i}{2}\bar\varkappa\bar\eta_\bk + \frac{1}{4}\varkappa \bar \varkappa \bar r_\bk
%= \SUM \, \bar \rho^{\rm even}_i \,\cdot\, \bar \rho^{\rm odd}_i\,,
\eea
where $\varkappa$ is the odd momentum (the Fourier image of $\theta$). We emphasize that bars on top of the operators indicate the LLL projection. Note that $\bar \eta_\bk^\dag = \bar \eta^\star_{-\bk} $ since $\bar \eta_\bk$ is a complex Grassmann operator.

Components of the superdensity operator form a non-trivial superalgebra given by the following relations together with \eqref{eq:GMP}
\bea\la{eq:GMR1}
&&[\bar \rho_\bk, \bar \eta_\bq] = f(\bk,\bq)\bar \eta_{\bk +\bq}\,, \quad [\bar \rho_\bk, \bar \eta^\star_\bq] = f(\bk,\bq)\bar \eta^\star_{\bk +\bq}\,,
\\\la{eq:GMR2}
&&\{\bar \eta_\bk, \bar \eta^\star_\bq\} = f(\bk,\bq)\bar r_{\bk +\bq}\,,\quad [\bar r_\bk, \bar \rho_\bq] = f(\bk,\bq) \bar r_{\bk +\bq}\,,
\\\la{eq:GMR3}
&& \{\bar \eta_\bk, \bar \eta_\bq\} = \{\bar \eta^\star_\bk, \bar \eta^\star_\bq\} = [\bar r_\bk,\bar r_\bq] = 0\,,
\eea 
where $f(\bk,\bq)$ is given by Eq.\eqref{eq:GMP}.
%and we have introduced a notation
%\be
%\bar \rho^{\rm even}_i = e^{-i\bar k \partial_i} e^{-\frac{i}{2} k z_i}\,,\quad \bar \rho^{\rm odd}_i=e^{-i\bar \varkappa \nabla_i} e^{-\frac{i}{2} \varkappa \theta_i}\,.
%\ee
%\be
% \left[\bar \rho_{k,\varkappa}, \bar \rho_{q,\gamma} \right] =  2i e^{\frac{\mathbf k \cdot \mathbf q - i \mathbf \varkappa \times \mathbf \gamma}{2}} \sin \frac{\mathbf k \times \mathbf q - i \mathbf \varkappa \cdot \mathbf \gamma}{2}\bar\rho_{\bk+\bq,\varkappa+\gamma}
%\ee
%At zero odd momentum $\bar \rho_{k,0}$ satisfies the usual GMP algebra. 

%%%%%%%%%%
\paragraph{Collective modes.---}
%%%%%%%%%%
The collective modes (both bosonic and fermionic) on top of the Moore-Read state are  given by a single expression
\be\la{eq:sGMP}
\Psi_{\bk,\varkappa} =\int \left[ \prod_{i=1}^{\mathrm{N}_{\rm el}}d\theta_{i}\right]  \bar \rho_{\bk,\varkappa} \Psi_{sL}\,.
\ee
To get some insight into the structure of $\Psi_{\bk,\varkappa}$ we first
assume that $\mathrm{N}_{\rm el}$ is even. Then the odd part of
$\Psi_{\bk,\varkappa}$ vanishes, because the superconformal block cannot involve
an odd number of fermions. Setting $\varkappa=\bar\varkappa=0$ we get
\be
\Psi_{GMP}(\mathbf{k})\equiv\Psi_{\bk,\varkappa=0} = \bar \rho_\bk \Psi_{MR}\,,
\ee
where $\bar \rho_\bk \Psi_{MR}$ is the standard GMP mode on top of the MR state. 

When $\mathrm{N_{\rm el}}$ is odd, the $\theta$-integral of $\Psi_{sL}$ itself vanishes identically since we are short of one $\theta$ (the number of $\theta$'s in the wavefunction is even by construction). However the integral in \eqref{eq:sGMP} does not vanish since $\bar \rho_{\bk,\varkappa}$ can contribute an additional $\theta$ and ensures that the integral does not vanish at the leading order in $\varkappa$. In fact, it is the odd operator $\bar \eta_\bk$ that creates the mode in the superspace. Since $\bar \eta_\bk$ is a superpartner of $\bar \rho_\bk$, the two collective modes are superpartners.
\begin{widetext}
Evaluating the integral over $\theta$ and setting $\bar\varkappa = 0$ we find the following trial state
\bea\la{eq:NF}
&&\Psi_{\bk,\varkappa} = -\frac{i}{2} \varkappa   \sum_{j=1}^{\mathrm N_{\rm el}}  (-1)^{j+1}f_{j} \Big[e^{-i\bar k \partial_j} e^{-\frac{i}{2} k z_j}\Delta^2 \Big]e^{- \SUM\frac{|z_i|^2}{4\ell^2}}\equiv \varkappa \Psi_{\rm NF}(\bk)\,,
\\
&&f_{j} = \mathcal A \left[ \frac{1}{z_1-z_2}\cdot \ldots \cdot \frac{1}{z_{j-2}-z_{j-1}} \frac{1}{z_{j+1}-z_{j+2}}\cdot \ldots \cdot \frac{1}{z_{\sst\rm N_{{\sst\rm el}}-1}-z_{{\sst\rm N_{\sst\rm el}}}} \right]\,,
\eea
where $\mathcal A$ stands for antisymmetrization. $f_{j}$ is the Pfaffian of a matrix $M_{ij} = \frac{1}{z_i-z_j}$, where $z_i$ are the coordinates of $\mathrm{N}_{\rm el}-1$ out of $\mathrm{N}_{\rm el}$ electrons, with $j$-th electron missing. The Vandermonde determinant involves all $\mathrm{N}_{\rm el}$ electrons.  The wavefunction \eqref{eq:NF} boosts the $j$-th electron and pairs the remaining electrons, the linear combination ensures that no electron is different from the others. The odd momentum $\varkappa$ serves as a book-keeping tool and can be discarded in the final expression.
\end{widetext}
The trial state $\Psi_{\rm NF}(\bk)$ is the first main result of the present Letter.  We will explore various properties of this state and its construction in the remainder of the letter.

First, we would like to make contact with the Jack polynomial construction of
\cite{yang2012model, 2013Yang}. At zero momentum $\bk=0$, the state $\Psi_{\rm
  NF}(\bk=0)$ coincides with the Pfaffian state at odd particle number (see
Ref.\cite{2013Yang} for an explicit formula). It also coincides with the highest
component of \eqref{superLaugh}. This state costs higher energy
\cite{bonderson2011numerical, moller2011neutral}, compared to the ground state
at \emph{even} particle number.  Eq.\ \eqref{eq:NF} tells us that at $\bk=0$ the wave function is a linear superposition of MR states with $\mathrm{N}_{\rm el}-1$ electrons and an extra charge $e$ quasihole placed at the position $z_i$. Complete anti-symmetrization over $z_i$'s then ensures that the wavefunction describes $\mathrm{N}_{\rm el}$ fermions. Next, we can analyze \eqref{eq:NF} at small $\bk$. To do so we expand $\Psi_{\rm NF}(\bk)$ in Taylor series in $\bk$, so that $\Psi_{\rm NF}(\bk)=\Psi_{\rm NF}(\bk=0) + \bar{k}^2 \Psi_{\rm 3/2} +\ldots$. 
Here, $\Psi_{\rm 3/2}$ is a polynomial in $z_i$, which up to normalization
coincides with the spin-$3/2$ state of Ref.\ \cite{yang2012model, 2013Yang}, given (up to the Gaussian factor) by
\be
\Psi_{3/2} =\Delta^2  \mathcal A \left[ \frac{1}{z_1-z_2}\cdot \ldots \cdot  \frac{1}{z_{2 \mathrm{N}-1}-z_{2\mathrm{N}}} \frac{1}{(z_1 - z_{2\mathrm{N} + 1})^2}  \right]\,. 
\ee
We conclude that $\Psi_{\rm NF}(\bk)$ is a trial state for a collective mode that agrees with this (Jack polynomial) construction at long wavelengths.

%%%%%%%%%%
\paragraph{Gap function.---}
%%%%%%%%%%

We now discuss the gap function of the neutral fermion mode, which is given by
\be
\Delta_{NF}(\bk) = \frac{\langle \bk_{\rm odd}| H |\bk_{\rm odd} \rangle}{\langle \bk_{\rm odd}| \bk_{\rm odd} \rangle} = \frac{\bar f_{\rm odd}(\bk)}{\bar \zeta(\bk)}\,,
\ee
where $\langle \bz|\bk_{\rm odd}\rangle = \Psi_{\rm NF}(\bk)\,$ and $\bar \zeta(\bk)$ is the norm of $|\bk_{\rm odd} \rangle$. Curiously, $ \bar \zeta(\bk)$ can be evaluated in the superspace as a two-point function of $\bar\eta_\bk$
\be
 \bar \zeta(\bk) = \int [d\theta] [d\bar \theta] [dz] [d\bar z] \Psi^*_{sL} \bar \eta^\star_{-\bk}\bar \eta_{\bk} \Psi_{sL} = \langle 0| \bar{\eta}^\star_{-\bk} \bar\eta_\bk|0\rangle\,,
\ee
where $[d\theta] = d\theta_1 d\theta_2\ldots d\theta_{\mathrm{N}_{\rm el}}$. 
Presently, no analytic results are available for the $\bar \zeta(\bk)$. 

The numerator, $\bar f_{\rm odd}(\bk)$ can be represented in terms of a  commutator and an \emph{anticommutator} as follows
\be
\bar f_{\rm odd}(\bk) = \frac{1}{2}\langle 0 | \{\bar \eta^\star_{-\bk}, [H,\eta_{\bk}]\}|0\rangle\,.
\ee
The anti-commutator appears because in order to use $k\rightarrow-k$ symmetry we need to exchange $\bar \eta$ and $\bar \eta^\star$, which anti-commute. Remarkably, the super algebra \eqref{eq:GMR1}-\eqref{eq:GMR3} can be utilized to express $f_{\rm odd}(\bk)$ in terms of the two-point functions (similar to the GMP mode)
\bea
&&\bar f_{\rm odd}(\bk) = 4\int d^2\bq V_\bq \left(\sin\left(\frac{\bk\times\bq}{2}\right)\right)^2F_{\rm odd}(\bk,\bq)\,,
\\
&&F_{\rm odd}(\bk,\bq) = e^{\bk \cdot \bq} \bar \zeta(\bk+\bq) + e^{- \frac{|\bk|^2}{2}} \bar \alpha(\bq)\,,
\eea
where we had to introduce another two-point function 
\be
\bar \alpha(\bq) = \langle 0| \bar \rho_{-\bq}\bar r_{\bq} |0\rangle\,.
\ee
No analytic results are yet available for $\bar\alpha(\bq)$. In principle one can compute any (Grassmann-even) two-point function of the generators of the superalgebra \eqref{eq:GMR1}-\eqref{eq:GMR3}. Only the two point function of bosonic densities --- the static structure factor --- has been studied before.
% We can utilize this fact to evaluate $\bar f_{\rm odd}(\bk)$ in terms of $\bar \zeta(\mathbf{k})$. We find
%\bea
%&&\bar f_{\rm odd}(\bk) = \frac{1}{2\mathrm N_{\rm el}}\int d^2 \bq~ 4\sin^2 \left( \bk\times \bq\right) V(\bq) \bar Z(\bq,\bk)\,, 
%\\\nonumber
%&&\bar Z(\bq,\bk) =  e^{\mathbf k\cdot \mathbf q}\bar \zeta(\bq+\bk)  - e^{-\mathbf k\cdot \mathbf q}\bar \zeta(\bq-\bk) -  2e^{\frac{|k|^2}{2}}\bar \zeta(\bq) \,.
%\eea
%The neutral fermion gap function, $\Delta_{NF}(\bk)$, is determined by the function $\bar \zeta(\bq)$ and interaction. Analytic properties of this function require further investigation.

%%%%%%%%%%
\paragraph{Collective modes from CFT.---}
%%%%%%%%%%

The superconformal blocks are usefully organized as multivariate functions on
the superplane $\mathbb R^{2|2}$.  One can cast the $\varkappa$
dependence as the parameter for a supersymmetry transformation of the background charge.  The superfield~\eqref{eq:charge superfield} combines the quasihole and electron operators.  The Grassmann parity of the background charge then correlates with that of the product of charged operator insertions.  

Infinitesimal superconformal transformations 
are parametrized by infinitesimal vector superfields
\be
\cV^z=v^z(z)+\theta\,v^\theta(z)
\ee
that shift the superspace coordinates via
%are generated by the differential operators
(see for instance~\cite{Friedan:1986rx,Polchinski:1998rr})
\be
\delta z = v^z(z) - i\theta v^\theta(z)
~~,~~~~
\delta \theta = -iv^\theta(z) +{\textstyle \frac12}\theta \, \partial_z v^z(z)
\ee
%\be
%L_n = -z^{n+1} \partial_z +  \frac {n+1}2 z^n \theta\partial_\theta\,,~~~~
%G_{n+\half} = z^{n+1}(\partial_\theta - \theta \partial_z).
%\ee
Acting on fields, these transformations are generated by the stress tensor supercurrent 
$\cT_{z\theta} = G_{z\theta}(z)+ \theta\, T_{zz}(z)$ via
\be
\cT_\cV\cdot \cO = \frac{1}{2\pi i} \oint_{\cC} dz d\theta \,\cV^z \cT_{z\theta}(z,\theta)\, \cO
\ee
with the contour $\cC$ surrounding $\cO$.
A superfield $\cO = \cO_0(z)+\theta\, \cO_1(z)$ of conformal weight $h$ transforms as
\begin{align}
\delta \cO_0 &= -\bigl[v^z\partial_z +h\, (\partial_z v^z)\bigr] \cO_0 - v^\theta \cO_1
\\
\delta \cO_1 &= -\bigl[v^z\partial_z +(h\!+\!{\textstyle \frac12}) (\partial_z v^z)\bigr] \cO_0
- \bigl[ v^\theta \partial+2h(\partial_z v^\theta)  \bigr] \cO_0
\nonumber
\end{align}
%In particular, for a superfield $\cO(z,\theta) = \cO_0(z)+\theta \cO_1(z)$ of conformal weight $h$ one has
%\footnote{This is a bit of an abuse of notation -- what we really mean is that the superdifferential operators act as Lie derivatives on the superfield, keeping in mind its tensorial nature, and then one evaluates the components.}
Expanding in mode operators 
%$G_n, L_n$ which are the operator coefficients of a Laurent series expansion
$G_{z\theta} = \sum_n \frac12 G_n z^{-n-3/2}$, $T_{zz} = \sum_n L_n z^{-n-2}$ and similarly for $v^z,v^\theta$, one has
\begin{align}
G_{\frac12}\cdot\cO_1 &= 2h\, \cO_0\,,~~~~ 
G_{-\half}\cdot\cO_0 = \cO_1   \\
G_{\half}\cdot\cO_0 &= 0\,,~~~~
G_{-\half}\cdot \cO_1 = L_{-1}\cdot \cO_0 = \partial_z\cO_0\,,
\nonumber
\end{align}
where the Laurent series is taken about $\cO$.
The operators $L_{\pm 1}, L_0, G_{\pm \frac{1}{2}}$ form a closed superalgebra $OSp(2|1)$.

We start by considering the CFT interpretation of the traditional GMP mode. The
long-wave expansion of the magnetoroton mode takes form $\Psi_{\rm
  GMP}(\mathbf{k}) =\Psi_0(z) + \bar k^2 \Psi_2(z)+ \ldots$. The polynomial
$\Psi_2(z)$ is the spin-$2$ part of the GMP mode. It coincides with spin-$2$
wavefunction $\Psi_2$ defined in Ref.\ \cite{yang2012model, 2013Yang}. $\Psi_2(z)$ can be also expressed in terms of the Virasoro generators inserted in the conformal block representation of the trial state (\emph{e.g.} Moore-Read state)
\be
\Psi_2(z) = \SUM (L_{-1}^{~2})_i \cdot \Psi_{\it MR} ~,
\ee
where $(L_{-1})_i$ is the Virasoro generator acting on $i$-th insertion.

Next, we turn to the fermionic collective mode. The state of an odd number of electrons, $\Psi_{\rm NF}(\mathbf{k}=0)$ is given by  
\be
\Psi_{\rm NF}(\mathbf{k}=0)=\int   \Biggl[\prod_{j=1}^{\mathrm{N}_{\rm el}}d\theta_{j}\Biggr] \SUM(G_{\half})_i \cdot \Psi_{\it sL}\,,
\ee
while the spin-3/2 state (which arises at the order
$\varkappa \bar k^2$) is given by
\be
\Psi_{\frac{3}{2}}(z) = \int \Biggl[\prod_{j=1}^{\mathrm{N}_{\rm el}}d\theta_{j}\Biggr] \SUM (L_{-1}^2 G_{\half})_i \cdot \Psi_{\it sL} 
\ee
(note that $(L_{-1}^{~2} G_{\half})_i=(L_{-1} G_{-\half})_i$ when acting on $\Psi_{\it sL}$).
We regard these results as an indication of the utility of (super)conformal methods in the analysis of collective excitations of trial wavefunctions for fractional quantum Hall states.

%%%%%%%%%%
\paragraph{Conclusions.---}
%%%%%%%%%%

We have investigated the collective neutral excitations in the Moore-Read phase, which is the candidate description for the $\nu=\frac{5}{2}$ observed plateau. It was shown that the two branches of collective excitations, at least at small momentum, can be naturally constructed in the auxiliary superspace formalism, which unifies the bosonic Laughlin and fermionic Moore-Read states into a single function, defined on a superspace. We have related our construction to previous work on the collective modes and found agreement, whenever such a comparison is possible. The gap function of the neutral fermion mode was expressed in terms of the norm of the neutral fermion state at finite momentum. Finally, we have shown that the trial states for both types of collective excitations can be constructed by inserting $OSp(2|1)$ (super-)Virasoro generators in the CFT representation of the trial states.

Our work suggests many new directions. First, it may be possible to extend the recently proposed effective bimetric theory of the collective spin-2 mode \cite{gromov2017Bimetric, nguyen2018fractional, liu2018geometric} to include the spin-$\frac{3}{2}$ fermionic mode. If the splitting between these modes is sufficiently small relative to the gap the effective theory should have an emergent, albeit softly broken supersymmetry. Second, it would be very interesting to revisit the plasma map for the MR state \cite{gurarie1997plasma,read2009non,bonderson2011plasma}; the plasma corresponding to \eqref{eq:sL} ``lives'' in the superspace and seems to have a simple structure. Third, the CFT approach may be useful to study the collective modes on top of the Read-Rezayi \cite{Read1999} states, which may support a richer collection of neutral modes. Finally, it is concievable that superspace techniques such as \cite{alvarez1992superloop, ciosmak2016super} can be utilized to directly compute $\bar s(\mathbf{k})$, $\bar \zeta(\mathbf{k})$ and $\bar \alpha(\bq)$ for the Moore-Read state, neither of which is known analytically via a direct computation.

%%%%%%%%%%
\paragraph{Acknowledgments.---}
%%%%%%%%%%

A.G. was supported by a Kadanoff Center fellowship and by the University of Chicago Materials Research Science and Engineering
  Center, which is funded by the National Science Foundation under award number
  DMR-1420709 and A.G. Quantum Materials program at
LBNL, funded by the US Department of Energy under
Contract No. DE-AC02-05CH11231.  EJM is supported in part by DOE grant DE-SC0009924.
  SR is supported by the National Science Foundation under
  award number DMR-1455296, and by 
 a Simons Investigator Grant from the Simons Foundation.

%%%%%%%%%%
%%%%%%%%%%

\bibliography{Bibliography}

\newpage

\end{document}